\documentstyle[12pt,epsfig]{article}

\def\pa{\partial}
\def\na{\nabla}
\def\nn{\nonumber}
\def\ep{{\rm e}^{-\phi}}

\def\sg{\sqrt{-g}}
\def\ga{\gamma}

\def\si{\sigma}

\def\de{\delta}
\def\ka{\kappa}
\def\al{\alpha}
\def\be{\beta}

\def\la{\lambda}
\def\La{\Lambda}
\def\eps{\epsilon}  
  
\def\sgn{{\rm sgn}}
 
\def\d{{\rm d}}

\begin{document}
\thispagestyle{empty}
\begin{flushright}
hep-th/0206254
\end{flushright}
\vspace{2cm}
\begin{center}
{\large\bf New brane solutions in higher order gravity}\\
\vspace{8mm}
{\bf \fbox{Adam Jak\'obek${}^a$}$\,$, 
Krzysztof A. Meissner${}^{a,}$\footnote{
\ \ E-mail: Krzysztof.Meissner@fuw.edu.pl}, 
and Marek Olechowski${}^{a,b,}$\footnote{
\ \ E-mail: olech@th.physik.uni-bonn.de}
} 
\\ 
\vspace{4mm}
{\it
${}^a$Institute of Theoretical Physics, Warsaw University\\ 
Ho\.za 69, 00-681 Warsaw, Poland\\
\vspace{4mm} 
${}^b$Physikalisches Institut der Universit\"at Bonn\\ 
Nussallee 12, 53115 Bonn, Germany
}
\end{center}
 
\vspace{1cm}
\begin{abstract}
We consider the higher order gravity with dilaton and with the leading
string theory corrections taken into account. The domain wall type
solutions are investigated for arbitrary number of space--time
dimensions. The explicit formulae for the fixed points and asymptotic
behavior of generic solutions are given.  
We analyze and classify solutions with finite effective gravitational   
constant. There is a class of such solutions which
have no singularities. 
We discuss in detail the relation between fine tuning and self tuning
and clarify in which sense our solutions are fine--tuning free. 
The stability of such solutions is also discussed.
\end{abstract}

\newpage
\section{Introduction}

Theory of closed strings incorporates gravity in a natural way
since the spin-2 field is always present in its massless spectrum.
{}From this point of view theory of gravity is an effective theory
of the graviton when the energies are smaller than the mass of the
string first excited level. String theory contains many other
massless fields and in an effective action one expects the
presence of all powers of the fields -- indeed the calculations
\cite{mt} of the string amplitudes involving the graviton,
dilaton and the antisymmetric tensor confirm that the string
effective action contains higher order terms.
The expansion parameter in string theory $\al'$ is dimensionful and
connected to 
the fundamental string length $l_s$ by $\al'=l_s^{-2}$. Since the
effective action was calculated in \cite{mt} (to order $\al'^2$)
by comparison with the string amplitudes on-shell, we cannot
distinguish in this way actions differing by lower order equations
of motion. However these actions from the field theory point of
view are different and the equations of motion have different
solutions. Since we don't know string amplitudes off-shell we have
to employ other methods to try to guess the true effective action
-- one possible idea is to look for an action with 
some additional, string motivated symmetry. In \cite{kam} it was
proven that one of the actions for gravity coupled to the dilaton
and the antisymmetric tensor at order $\al'^2$ exhibits an
$O(d,d)$ symmetry characteristic for string effective actions
\cite{mv}. Therefore in the present paper we use the form of the
action 
obtained in \cite{kam} and analyze the presence of the domain wall
type solutions of the equations of motion.

It was shown in \cite{KMO} that in the presence of arbitrary
number of Euler densities in the Lagrangian (without the
dilaton), there always exists a domain wall solution of the
Randall--Sundrum type \cite{RS2}.  In the present paper we analyze
the gravity theory coupled to the dilaton with interaction terms up to
the fourth order in the derivatives. 
Properties of the generic domain wall type solutions are investigated.
We are interested mainly in those solutions which allow for 
localized gravity with finite effective gravitational constant
after integrating over the coordinate perpendicular to the brane.
We find that both with and without the
cosmological constant the presence of the dilaton allows for
solutions of the Randall--Sundrum type \cite{RS2} 
as well as solutions with singularities of the type
described in \cite{KSS}. We find also some new cases specific to
higher order gravity. 
It turns out that some of the solutions without singularities have 
finite effective gravitational constant without
fine tuning of the parameters of the Lagrangian. The problem of
the absence of fine tuning and potential self  
tuning (for the discussion in the case of the lowest order gravity see
\cite{FLLN}) 
of these solutions is addressed in great detail. 
We explain the relation between our fine--tuning free solutions and
some no--go theorems \cite{NoGo1,NoGo2} derived for theories similar
to the one consider in the present paper.
Recently one similar solution for a 5--dimensional theory has been
found in ref.\ \cite{BCDD}. However the higher order terms used 
in \cite{BCDD} are as considered in ref.\ \cite{MR1} and 
differ from those adopted in the preset work (we 
explain the reason for this difference in the next section). Moreover,
we present more solutions of this type and our analysis can be applied
to theories in arbitrary dimensions. We discuss also, contrary to
\cite{BCDD}, the stability of such fine tuning free solutions.
Some exact solutions in higher order gravity have been recently
discussed in \cite{MR2}.

The paper is organized as follows:
In section 2 we present the Lagrangian and derive the equations of
motion 
in general and in the application to the brane metric. In section 3 we 
explicitly find fixed points and asymptotic behavior of
solutions. The  
number of the fixed points and the  behavior of asymptotics depends 
very much on the value of the bulk cosmological constant. 
In section 4 we 
analyze and classify solutions with finite effective gravitational 
constant - the finiteness is due either to the compactness of the 
transverse dimension or to the decreasing warp factor like in the
Randall-Sundrum scenario. 
Section 5 is devoted to the 
detailed discussion of the equations of motion at the brane. 
In section 6 we 
discuss the problem of fine tuning of the solutions with
finite effective gravitational constant and  
clarify in which sense our solutions avoid fine tuning.
The relation of our results to some no--go theorems \cite{NoGo1,NoGo2}
is also explained. In section 7 we present the conclusions.

\section{Higher order action and equations of motion}

In theories with gravity coupled to the dilaton one has to decide
upon the so called frame (i.e. choice of metric) since one can
redefine the metric by a (non-vanishing) factor depending on the
dilaton. There are usually two frames used: the string frame (with
the factor $e^{-\phi}$ in front of the curvature scalar in the
string tree level effective action) and the Einstein frame
(without this factor). The metrics solving the equations of motion
can have quite different properties in the two frames (for a
discussion in the cosmological context see \cite{GV}). 
In the present Universe and with the
presently available energies the difference between the frames is
unimportant since the dilaton seems to be (by a still unknown
mechanism) constant both in time and in space, but both in the
early Universe or, as discussed in this paper, close to a brane,
the difference can be quite substantial. It is generally argued
that although the physical conclusions should not depend on the
redefinition of fields, it is much easier to draw these
conclusions in the string frame than in the Einstein frame and the
''translation'' from one to the other is not always evident. 
Therefore we will use the string frame in this paper.
It should be pointed out that the higher order terms considered in
refs.\ \cite{BCDD,MR2} are different from those used in the present
paper even after changing the frame.

The string motivated action we consider reads \cite{kam} 
(note a difference by a
factor of 2 in normalization of the dilaton $\phi$): 
\begin{equation}
S=S_0+S_1+S_B=\int \d^Dx\left(L_0+\al L_1+L_B\right)
\,,
\label{action}
\end{equation}
where $\al=\al'\la_0$ and
\begin{eqnarray}
L_0
&\!\!\!=\!\!\!&
\frac1{2\ka^2}
\sg\ep\left[-2\La+R+\pa_\mu\phi\pa^\mu\phi\right]
\label{L0}
\,,
\\[6pt]
L_1
&\!\!\!=\!\!\!& 
\frac1{2\ka^2}
\sg\ep 
\left[-R_{GB}^2 
-2\pa_\mu\phi\pa^\mu\phi\Box\phi
\right.
\nn\\[6pt]
&&\qquad\qquad\qquad\left.
+4\left(R^{\mu\nu}-\frac12 g^{\mu\nu}R\right)
\pa_\mu\phi\pa_\nu\phi 
+(\pa_\mu\phi\pa^\mu\phi)^2\right]
\label{L1}
,
\\[6pt]
L_B
&\!\!\!=\!\!\!&
\frac1{2\ka^2}
\sqrt{-\tilde{g}}V(\phi)\de(y)
\label{LB}
\,.
\end{eqnarray} 
$R_{GB}$ is the Gauss--Bonnet term
\begin{equation} 
R_{GB}^2=R^{\mu\nu\rho\si}R_{\mu\nu\rho\si} 
-4R^{\mu\nu}R_{\mu\nu}+R^2
\,.
\label{GB}
\end{equation}
The space--time has arbitrary dimension $D$. The metric in eq.\
(\ref{LB}) is the induced metric on the brane
\begin{equation}
\tilde{g}_{\al\be}=g_{\mu\nu}\de^\mu_\al\de^\nu_\be
\,.
\end{equation}
The indices have the following ranges: $\mu,\nu=1,\ldots,D$;
$\al,\be=1,\ldots,d$; where $d=D-1$. The coordinate perpendicular to
the brane we denote by $y=x^D$. We fixed the position of the brane at
$y=0$.

Let us now calculate the functional derivative of this
action (separately for each term in (\ref{action})) with
respect to variations of the metric and to variations of
the dilaton. We define 
\begin{eqnarray}
T_{\mu\nu}^{(I)}
&\!\!\!=\!\!\!&
2\ka^2\frac{{\rm e}^\phi}{\sg}\frac{\de}{\de g^{\mu\nu}}\int \d^Dx L_I
\,,
\\[6pt]
W^{(I)}
&\!\!\!=\!\!\!&
2\ka^2\frac{{\rm e}^\phi}{\sg}\frac{\de}{\de\phi}\int \d^Dx L_I
\,,
\end{eqnarray}
for $I=0,1,B$.
A rather tedious calculation gives for $T_{\mu\nu}$ 
\begin{eqnarray} 
T_{\mu\nu}^{(0)}
&\!\!\!=\!\!\!&
\La g_{\mu\nu}+ R_{\mu\nu}-\frac12
g_{\mu\nu}R+\frac12g_{\mu\nu}\pa_\si\phi\pa^\si\phi
+\na_\mu\pa_\nu\phi-(\Box\phi)g_{\mu\nu}
\,,
\label{T0}
\\[2pt]
T_{\mu\nu}^{(1)}
&\!\!\!=\!\!\!&
4R^{\ka\la}R_{\ka\mu\la\nu}
-2R_{\mu\si\ka\la}R_\nu{}^{\si\ka\la}
+4R_{\mu\ka}R^\ka{}_\nu
-2RR_{\mu\nu}
+\frac12 g_{\mu\nu}R_{GB}^2
\nn\\[2pt]
&&\!\!\!\!\!
+4(\na^\ka\pa^\la\phi) R_{\mu\ka\nu\la}
+3\pa_\ka\phi\pa^\ka\phi R_{\mu\nu}
+8(\na_\ka\pa_\mu\phi)R^\ka{}_\nu
-4(\Box\phi)R_{\mu\nu}
\nn\\[6pt]
&&\!\!\!\!\!
-4(\na_\ka\pa_\la\phi)R^{\ka\la}g_{\mu\nu}
-2(\pa_\ka\phi\pa^\ka\phi)R g_{\mu\nu}
-2(\na_\mu\pa_\nu\phi)R
+2(\Box\phi)R g_{\mu\nu}
\nn\\[6pt]
&&\!\!\!\!\!
-2\pa^\ka\phi\pa_\ka\phi\Box\phi g_{\mu\nu}
+2(\Box\phi)^2 g_{\mu\nu}
-2(\na^\ka\pa^\la\phi)(\na_\ka\pa_\la\phi) g_{\mu\nu}
+2\pa_\ka\phi\pa^\ka\phi\na_\mu\pa_\nu\phi
\nn\\[6pt]
&&\!\!\!\!\!
+4(\na_\mu\pa_\ka\phi)(\na_\nu\pa^\ka\phi)
-4(\na_\mu\pa_\nu\phi)\Box\phi
+\frac12(\pa^\ka\phi\pa_\ka\phi)^2 g_{\mu\nu}
\,,
\label{T1}
\\[2pt]
T_{\mu\nu}^{(B)}
&\!\!\!=\!\!\!&
-\frac{e^\phi}{2} \tilde{g}_{\al\beta}\de^\al_\mu
\de^\beta_\nu V(\phi)\de(y)
\,.
\label{TB}
\end{eqnarray} 
Calculation of the functional derivatives with respect to
$\phi$ is straightforward 
\begin{eqnarray} 
 W^{(0)}
&\!\!\!=\!\!\!&
2\La +\pa_\ka\phi\pa^\ka\phi
-2\Box\phi-R
\,,
\label{W0}
\\[6pt] 
W^{(1)}
&\!\!\!=\!\!\!&
R_{GB}^2
+4\left(R^{\ka\la}
-\frac12 Rg^{\ka\la}\right)(\pa_\ka\phi\pa_\la\phi
-2\na_\ka\pa_\la\phi)
+(\pa_\ka\phi\pa^\ka\phi)^2
\nn\\[2pt]
&&\!\!\!\!\!
+4(\Box\phi)^2 
-4(\na^\ka\pa^\la\phi)(\na_\ka\pa_\la\phi)
-4R^{\ka\la}\pa_\ka\phi\pa_\la\phi
\,,
\label{W1}
\\[6pt]
W^{(B)}
&\!\!\!=\!\!\!&
e^\phi V'(\phi)\de(y)
\,.
\label{WB}
\end{eqnarray}
The equations of motion read 
\begin{eqnarray}
T_{\mu\nu}^{(0)}+ 
\al T_{\mu\nu}^{(1)}
+T_{\mu\nu}^{(B)}=0 
\,,
\label{sumT}
\\[6pt]
 W^{(0)}+\al W^{(1)}+W^{(B)}=0 
\,.
\label{sumW} 
\end{eqnarray} 
Let us note that in the bulk, i.e  outside of branes and
singularities, we have the equality (up to boundary terms) that can be
easily checked
\begin{equation}
S=-\frac{1}{2\ka^2}\int \d^D x\sg\ep\left[
 W^{(0)}+\al W^{(1)}\right].
\end{equation}
It means that on--shell i.e. on the equations of motion the effective 
cosmological constant has no contribution from the bulk. On physical 
grounds, however, we are interested in the solutions for which the total 
(i.e. including branes) 
effective  cosmological constant vanishes on--shell. Anticipating the 
results of section 5, for fixed point solutions it happens for one form 
of the potential on the brane, namely  $V(\phi)=\la e^{-\phi}$ i.e. the 
potential suggested by string theory (for NS fields for which the bulk 
action is considered). For special solutions which are not exactly at the 
fixed points it happens for larger class of brane potentials. 
Then the boundary terms from the bulk cancel the brane contribution and we 
have 
\begin{equation} 
S=-\frac{1}{2\ka^2}\int \d^D x\sg\ep\left[
 W^{(0)}+\al W^{(1)}+W^{(B)}\right].
\end{equation}
where the integral is over the full space including brane positions. 
Since it is proportional to the equations of motion, it vanishes on--shell 
so the effective cosmological constant vanishes as well.

We now start with solving the equations in the bulk and later
we will include $T_{\mu\nu}^{(B)}$ and $W^{(B)}$.
The line element of the domain wall background is equal to 
\begin{equation}
\d s^2=e^{2A(y)} \eta_{\al\beta} \d x^\al \d x^\beta + \d y^2
\,.
\label{ds2}
\end{equation}
The three equations of motion -- $\al\beta$
and $yy$ components of (\ref{sumT}), and (\ref{sumW}) --
read respectively:
\begin{eqnarray} 
0
&\!\!\!=\!\!\!&
\La +(d-1)A''-\phi''+\frac12
d(d-1)(A')^2-(d-1)A'\phi'+\frac12(\phi')^2
\nn\\[2pt]
&&\!\!\!\!\!
+\al\left[
2(d-1)(d-2)(d-3)A''(A')^2 
-4(d-1)(d-2)A''A'\phi' 
+2(d-1)A''(\phi')^2
\right.
\nn\\[6pt]
&&\!\!\!\!\!
\left.\ \ \ \ \
-2(d-1)(d-2)\phi''(A')^2
+4(d-1)\phi''\phi'A'
-2\phi''(\phi')^2
\right.
\nn\\[2pt]
&&\!\!\!\!\!
\left.\ \ \ \ \
+\frac12d(d-1)(d-2)(d-3)(A')^4 
-2(d-1)^2(d-2)(A')^3\phi'
\right.
\nn\\[2pt]
&&\!\!\!\!\!
\left.\ \ \ \ \
+(3d-4)(d-1)(A')^2(\phi')^2
-2(d-1)A'(\phi')^3+\frac12(\phi')^4\right], 
\label{EoMmn} 
\\[2pt]
0
&\!\!\!=\!\!\!&
\La+\frac12
d(d-1)(A')^2-dA'\phi'+\frac12(\phi')^2
\nn\\ [2pt]
&&\!\!\!\!\!
+\al\left[ 
\frac12d(d-1)(d-2)(d-3)(A')^4 
-2d(d-1)(d-2)(A')^3\phi'
\right.
\nn\\[2pt]
&&\!\!\!\!\!
\left.\ \ \ \ \
+3d(d-1)(A')^2(\phi')^2
-2dA'(\phi')^3
+\frac12(\phi')^4\right], 
\label{EoMyy} 
\\[2pt]
0
&\!\!\!=\!\!\!&
\La+dA''-\phi''+\frac12
d(d+1)(A')^2-dA'\phi'+\frac12(\phi')^2
\nn\\[2pt]
&&\!\!\!\!\!
+\al\left[
2d(d-1)(d-2)A''(A')^2 
-4d(d-1)A''A'\phi'
+2dA''(\phi')^2
\right.
\nn\\[6pt]
&&\!\!\!\!\!
\left.\ \ \ \ \
-2d(d-1)\phi''(A')^2
+4d\phi''\phi'A'
-2\phi''(\phi')^2
+\frac12d(d+1)(d-1)(d-2)(A')^4 
\right.
\nn\\[2pt]
&&\!\!\!\!\!\ \ \ \ \
\left.
-2d^2(d-1)(A')^3\phi'+d(3d-1)(A')^2(\phi')^2
-2dA'(\phi')^3+\frac12(\phi')^4\right].
\label{EoMdil} 
\end{eqnarray} 
These equations have two obvious symmetries 
\begin{equation} 
y\to y+y_0\,,\ \ \ A\to A+A_0\,,\ \ \
\phi\to \phi+\phi_0\,, 
\label{symm1} 
\end{equation} 
and 
\begin{equation} 
y\to -y\,,\ \ \ A'\to -A'\,,\ \ \ \phi'\to -\phi'\,, 
\label{symm2} 
\end{equation} 
which can be used to obtain a whole class of solutions starting 
from each particular one.

Only two of the equations (\ref{EoMmn})--(\ref{EoMdil}) are independent
in the bulk. One of the combinations of those equations is
particularly simple. Multiplying the difference of (\ref{EoMmn}) and
(\ref{EoMdil}) by $\exp(\phi-dA)$ we get an expression which is a
total derivative with respect to $y$. Integrating it we obtain
\begin{equation}
e^{dA-\phi}
\left\{A'\left[
1+2\al\left((d-1)(d-2)A'^2-2(d-1)A'\phi'+\phi'^2\right)
\right]\right\}={\rm const}
\,.
\label{EoMconst}
\end{equation}
This equation will prove to be very useful when analyzing the fixed
points of the equations of motion.

\section{Fixed points and asymptotic behavior of solutions}
\label{FixedPoints}

Let us find the fixed points of the equations of motion 
(\ref{EoMmn})--(\ref{EoMdil}) 
(i.e. values of $A'$ and $\phi'$ for which 
$A''=\phi''=0$). We denote 
\begin{equation} 
A'=\frac{a_1}{\sqrt{\al}}
\,,
\qquad
\phi'=\frac{b_1}{\sqrt{\al}}
\,,
\label{FP}
\end{equation} 
where $a_1$, $b_1$ are constants. Substituting constant derivatives
$A'$ and $\phi'$ into eq.\ (\ref{EoMconst}) we see that it can
be fulfilled only in two ways: either $\phi'=dA'$ (only then the
exponential factor of (\ref{EoMconst}) is constant) or the expression
in the curly bracket in (\ref{EoMconst}) must vanish. It is not
difficult to check that only the second possibility is compatible with
eqs.\ (\ref{EoMmn}) and (\ref{EoMyy}). Thus, every fixed point of
eqs.\ (\ref{EoMmn})--(\ref{EoMdil}) must satisfy
\begin{equation}
a_1\left[1+2(d-1)(d-2)a_1^2-4(d-1)a_1b_1+2b_1^2\right]=0
\,.
\label{fixed}
\end{equation}
There are two types of fixed points corresponding to two ways in which
the above equation can be fulfilled. The first type has vanishing
$a_1$. The constant $b_1$ can be then calculated from
eq. (\ref{EoMyy}). The result reads:
\begin{equation}
a_1=0\,,
\qquad 
b_1=\pm\si\,,
\label{FP0}
\end{equation}
where
\begin{equation}
\si=\sqrt{\frac{\sqrt{1-8\al\La}-1}{2}}
\,.
\label{sigma}
\end{equation}
The number of such fixed points depends on the sign of the bulk
cosmological constant: there are two for $\La<0$, one for $\La=0$ and
none for $\La>0$. The solution for the vanishing bulk cosmological
constant has $a_1=b_1=0$ and is just the 5--dimensional Minkowski
space--time.

For the fixed points with nonzero $a_1$ the expression in the square
bracket in eq.\ (\ref{fixed}) must vanish. This gives a relation
between $a_1$ and $b_1$ which together with eq.\ (\ref{EoMyy})
gives the second type of the fixed point:
\begin{equation}
a_1=\eps_a\ga
\,,\qquad
b_1=\eps_a(d-1)\ga+\eps_b\sqrt{(d-1)\ga^2-\frac12}
\,,
\label{FPnon0}
\end{equation}
where $\eps_a\,,\eps_b=\pm1$ and $\ga$ can take one of the two
possible values, $\ga_+$ or $\ga_-$, given by
\begin{equation} 
\ga_\pm=\sqrt{\frac{1}{2(d-2)}\left(1\pm\sqrt{\frac{d}{2(d-1)}
+\frac{4\al\La(d-2)}{d-1}} 
\right)}
\,.
\label{gamma}
\end{equation} 
The requirement that the parameters $a_1$ and $b_1$ are
real gives two possibilities:
\begin{equation}
\ga=\ga_+  \qquad\qquad\ \, {\rm for}  \quad \La\ge\La_0\,,
\label{Lambdaplus}
\end{equation}
or
\begin{equation}
\ga=\ga_-  \qquad\qquad\ \, {\rm for}  \quad 
\La_1\ge\La\ge\La_0\,,  
\label{Lambdaminus}
\end{equation}
where
\begin{eqnarray}
\La_0
&\!\!=\!\!&
-\frac{d}{8\al(d-2)}
\,,\\[6pt]
\La_1
&\!\!=\!\!&
\La_0+[4\al(d-1)(d-2)]^{-1}
\,.
\label{Lambda01}
\end{eqnarray}
If solutions for $\ga=\ga_+$ ($\ga=\ga_-$) exist, there are
four of them corresponding to four combinations of 
signs $\eps_a$ and $\eps_b$ in eq.\ (\ref{FPnon0}).
The solutions for $\ga=\ga_-$ exist only for a rather small range 
(\ref{Lambdaminus}) of values of the bulk cosmological constant.

Adding all the above types of solutions  we can
see that the eqs.\ (\ref{EoMmn})--(\ref{EoMdil}) have the following
number of fixed points:
2 for $\La<\La_0$; 4 for $\La=\La_0$; 10 for $\La_0<\La<\La_1$; 8 for 
$\La=\La_1$; 6 for $\La_1<\La<0$; 5 for $\La=0$; 4 for $\La>0$.

Let us now analyze the asymptotic behavior of $A'$ and $\phi'$. 
There are different types of behavior
related to each of the allowed fixed points. One can find also 
asymptotic behavior for $A'$ and $\phi'$ going to infinity. The
results are as follows:
\begin{itemize} 
\item
There is one fixed point with $A'=0=\phi'$ if $\La=0$. 
The asymptotic behavior for $y\to \pm\infty$ is given by
\begin{equation} 
A'=\frac{1}{\sqrt{d}|y|}
\,,\qquad
\phi'=\frac{\sqrt{d}-\sgn(y)}{|y|} 
\,.
\label{as1}
\end{equation} 
\item
For the two fixed points with $A'=0$ for $\La<0$ and $y\to\pm\infty$
we have
\begin{equation}  
A'=a_2\exp(-\si |y|/\sqrt{\al})
\,,\qquad
\phi'=-\sgn(y)\frac{\si}{\sqrt{\al}}+da_2 \exp(-\si |y|/\sqrt{\al})
\,,
\label{as2}
\end{equation} 
where $\si$ is given by eq.\ (\ref{sigma}). 
\item 
If the bulk cosmological constant satisfies the condition
$\La_0<\La$ there are four fixed points 
$A'=\pm\ga/\sqrt{\al}$ with $\ga=\ga_+$ corresponding to
four combinations of signs $\eps_a$, $\eps_b$ in the expression for
$b_1$ given in eq.\ (\ref{FPnon0}). In addition if $\La_0<\La<\La_1$ 
there are four fixed points with $\ga=\ga_-$. Half of these fixed points
are approached for $y\to+\infty$ and the other half for $y\to-\infty$. 
Let us analyze solutions approaching the fixed points for $y\to\infty$. The
solutions can approach them either from above ($a_3>0$) or from below
($a_3<0$) and their asymptotic behavior is given by 
\begin{equation} 
A'=\eps_a\frac{\ga}{\sqrt{\al}}+a_3\exp(-cy/\sqrt{\al})
\,,\qquad
\phi'=\frac{b_1}{\sqrt{\al}}+b_3\exp(-cy/\sqrt{\al}) 
\,,
\label{as3}
\end{equation} 
where 
\begin{eqnarray} 
c&=&\eps_a\ga-\eps_b\sqrt{(d-1)\ga^2-\frac12}
\,,
\label{c}
\\[2pt]
b_3/a_3
&=&
d\left[1-
\left[2(d-1)\ga\left(\eps_a \eps_b\sqrt{(d-1)\ga^2-\frac12}+\ga\right)
\right]^{-1}
\right]^{-1}.
\label{b3a3}
\end{eqnarray} 
{}From the condition $c>0$ and the fact (cf. (\ref{gamma})) that
for $d>2$ 
\begin{equation} 
\sqrt{(d-1)\ga_+^2-\frac12}>\ga_+
\,,\qquad
\sqrt{(d-1)\ga_-^2-\frac12}<\ga_-
\,.
\label{gammapm}
\end{equation} 
we conclude that solutions with $\ga=\ga_+$ approach (as $y\to\infty$) 
the fixed point when $\eps_b=-1$ while those with $\ga=\ga_-$ when
$\eps_a=1$.

\item 
There are vertical asymptotes for each value of $y$. Denoting the 
location of the vertical asymptote as
$y_0$ one can check that $|A'|$ 
and $|\phi'|$ go to infinity for $y\to y_0$ as  
\begin{equation} 
A'=\frac{c_i}{y-y_0}
\,,\qquad 
\phi'=\frac{dc_i-3}{y-y_0} 
\,,
\label{as4}
\end{equation} 
where $c_i$ are four solutions (two positive and two
negative) of the equation 
\begin{equation} 
d(d-2)c_i^4+8dc_i^3-18dc_i^2+27=0 
\,.
\label{ci} 
\end{equation} 
$A'$ is positive for positive (negative) $c_i$ and $y>y_0$ ($y<y_0$).
The solutions with negative $A'$ are as usual related to those with
$A'>0$ by the symmetry (\ref{symm2}). Let us note that 
$c_i<3/2$ for $d>2$.
\end{itemize}

\begin{figure}[h]
\begin{center}
\epsfig{file=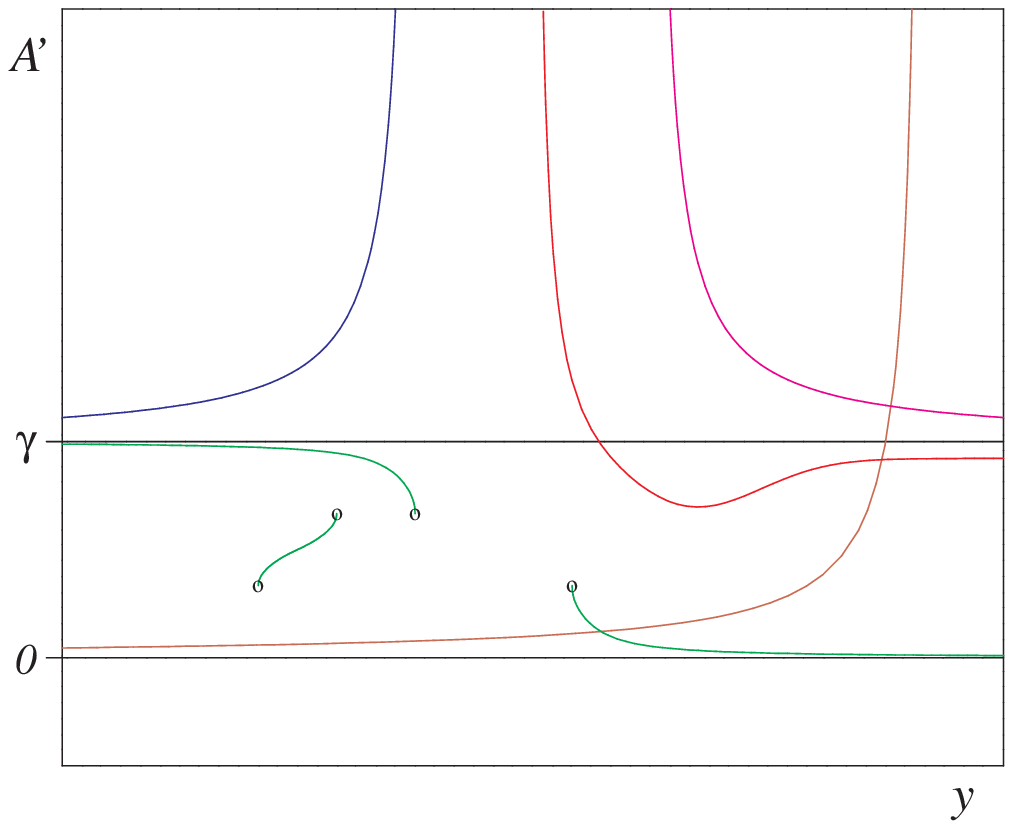,width=10cm}
\end{center}
\caption{\protect Typical solutions of the equations of motion
(\ref{EoMmn})--(\ref{EoMdil}) for $d<9$ and the bulk cosmological
constant in the range $\La_1<\La<0$. Only solutions with positive $A'$
are shown. The cases with negative $A'$ can be obtained by means of
symmetry (\ref{symm2}). From the symmetry (\ref{symm1}) it follows
that any of the solutions gives another solution when shifted by
arbitrary constant along the $y$ coordinate. The circles denote
singularities of type B (see discussion in section \ref{finiteG}) for
which the metric is non zero but the curvature tensor diverges.
}
\label{fig:d-}
\end{figure}

\begin{figure}[htbp]
\begin{center}
\epsfig{file=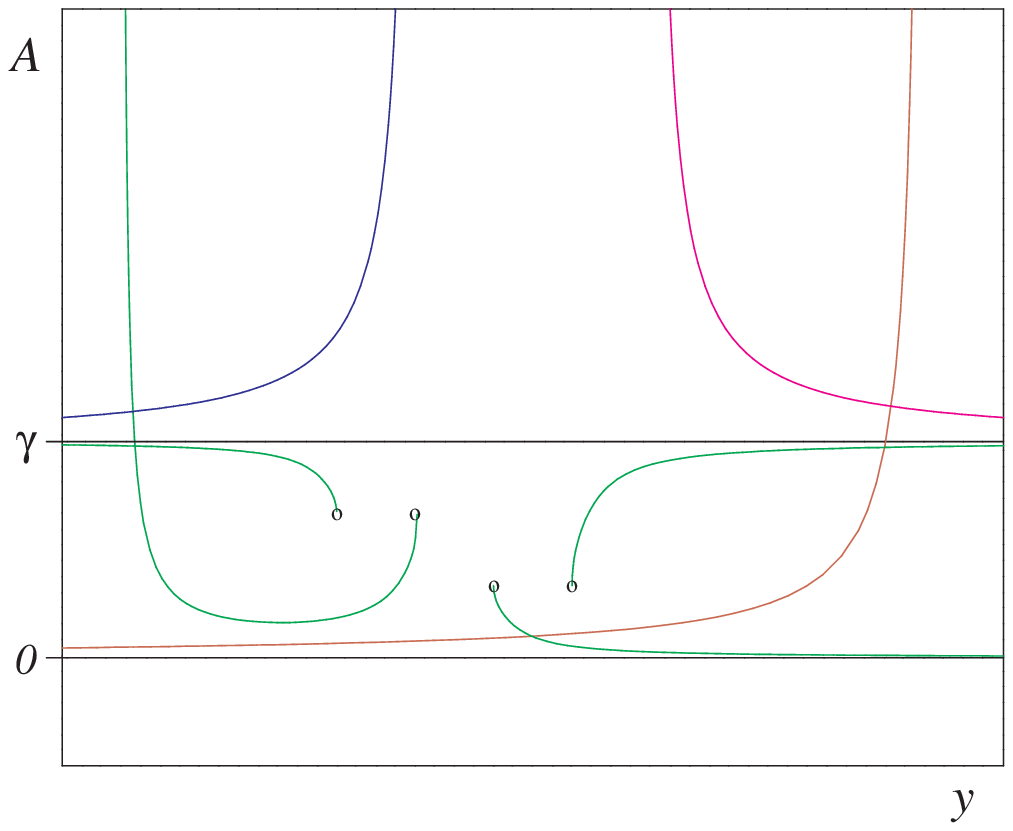,width=10cm}
\end{center}
\caption{\protect Same as figure 1 but for $d=9$}
\label{fig:d9}
\end{figure}

\begin{figure}[htbp]
\begin{center}
\epsfig{file=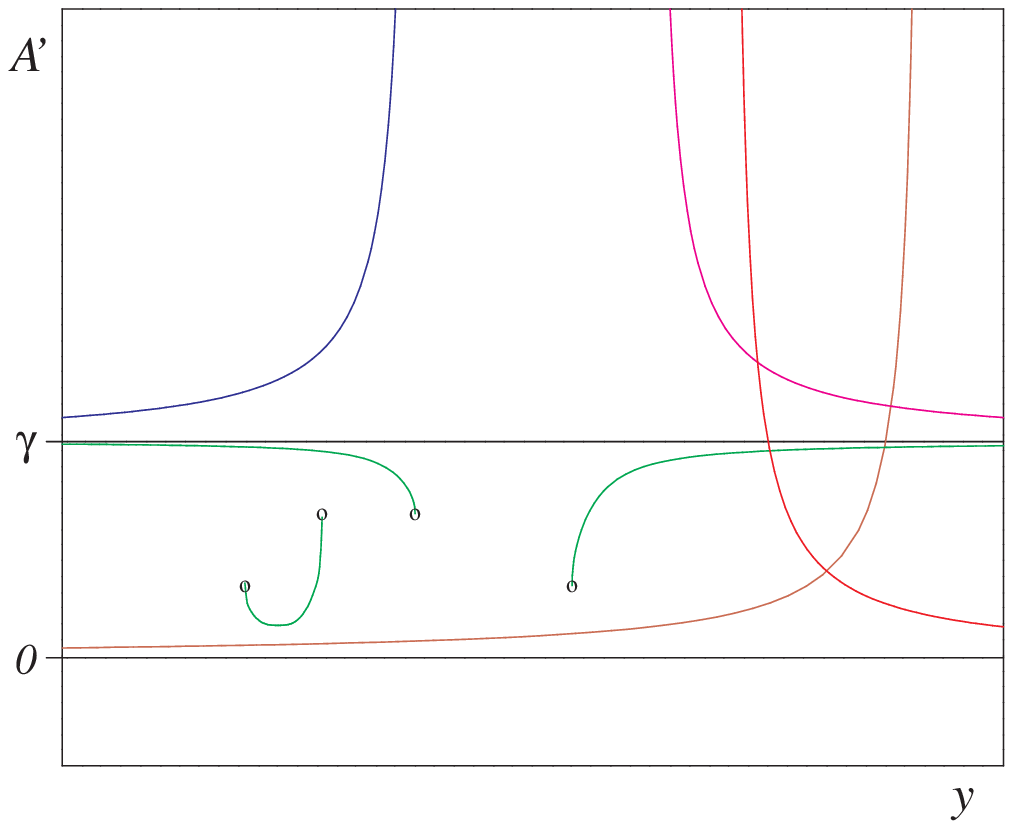,width=10cm}
\end{center}
\caption{\protect Same as figure 1 but for $d>9$}
\label{fig:d+}
\end{figure}

The above discussed fixed points and asymptotic behavior have been
obtained analytically from the equations of motion
(\ref{EoMmn})--(\ref{EoMdil}). Of course it is not possible to find
exact expressions for solutions which are not at the fixed
points. One has to use numerical methods to investigate such generic
solutions. On figures 1--3 we present some typical results. The
character of some of the solutions depends on the dimension of
the space--time and changes at $d=9$.

\section{Solutions with finite gravitational constant} 
\label{finiteG}

The solutions of the equations of motion discussed in the
previous  section may be used to construct different types of
domain wall models. Of course the most interesting
are those cases for which the lower  dimensional effective
gravitational constant is finite. The effective  lower
dimensional gravitational constant is obtained by
integrating  the metric over the direction perpendicular
to the wall. The are two  types of situations when that
constant is finite. First, the range of  the $y$
coordinate may be finite because the space ends at some 
singularities. This type of solutions was discussed for
the lowest  order string inspired dilaton gravity in 
\cite{KSS}.  Second, the range of $y$ may be infinite but with the 
warp  factor decreasing strongly enough to make the
integral finite.  This type of warped solutions is not
present in the lowest order  theory discussed in 
\cite{KSS} but appears in the gravity  theory without
dilaton at the lowest order \cite{RS2} and in higher 
orders \cite{KMO}.

In the higher order string inspired dilaton theory
investigated in  the present work both mechanisms to get a
finite effective  gravitational constant can be realized.
In fact we can define even three classes of such models because 
there are two different types of  singularities which can be
used to cut off the space.

Before we present all those classes of solutions a comment on the
effective gravitational constant is in order. 

In calculating the effective gravitational coupling 
we assume that the $D$-dimensional metric takes the form
\begin{equation}
\d s^2=e^{2A(y)} g^{(d)}_{\al\beta} \d x^\al \d x^\beta + \d y^2
\,.
\label{ds2comp}
\end{equation}
and
\begin{equation}
\phi=\phi(y)+\varphi(x^\al)
\,.
\label{phicomp}
\end{equation}
We then calculate $D$-dimensional quantities in terms of $d$ dimensional 
ones:
\begin{equation}
R_{\al\beta\ga\de}=e^{2A}R_{\al\beta\ga\de}^{(d)}
-e^{4A}A'^2
\left(g_{\al\ga}^{(d)}g_{\beta\de}^{(d)}
-g_{\al\de}^{(d)}g_{\beta\ga}^{(d)}\right)
\,.
\label{rabgd}
\end{equation} 
and
\begin{equation}
R_{\al y\beta y}=-e^{2A}g_{\al\beta}^{(d)}
\left(A'^2+A''\right)
\,.
\label{rayby}  
\end{equation}
Hence we have\footnote{Note that our formulae are different from the ones 
derived in  \cite{MR2} and used in \cite{BCDD}}
\begin{eqnarray}
R_{\al\beta}&=&R_{\al\beta}^{(d)}
-e^{2A}g_{\al\beta}^{(d)}\left(dA'^2+A''\right)
\nn\\
R_{yy}&=&-d\left(A'^2+A''\right)
\nn\\
R&=&e^{-2A}R^{(d)}-d\left((d+1)A'^2+2A''\right)
\nn\\
R_{GB}&=&e^{-4A}R_{GB}^{(d)}-2(d-2)e^{-2A}R^{(d)}
\left((d-1)A'^2+2A''\right)\nn\\
&&+d(d-1)(d-2)\left((d+1)A'^4+4A''A'^2\right)
\label{rcomp}
\end{eqnarray}
Then the result of the integration over $y$ of the action
(\ref{action}) gives the string frame $d$-dimensional action
\begin{equation}
\int d^d x\sqrt{-g^{(d)}}
e^{-\varphi(x)}\left(C_1+C_2R^{(d)}+\ldots\right)\,.
\label{compact}
\end{equation}

The $d$-dimensional cosmological constant is proportional to $C_1$ 
i.e. to the integral 
\begin{equation}
\int\d y \exp\left[dA(y)-\phi(y)\right]P_1(A',A'',\phi',\phi''),
\label{effL}   
\end{equation}
while the inverse of the $d$-dimensional gravitational constant
is proportional to $C_2$ i.e. to the integral
\begin{equation}
\int\d y \exp\left[(d-2)A(y)-\phi(y)\right]P_2(A',A'',\phi',\phi'').
\label{effG}
\end{equation}
where $P_1(A',A'',\phi',\phi'')$ and $P_2(A',A'',\phi',\phi'')$ are some 
polynomials. 
If these polynomials do not vanish on the solution for other reasons (as 
happens for the cosmological constant on-shell)
the behaviour of the exponential allows us to find whether a given solution 
has finite or infinite cosmological and gravitational constants. 
 
Let us discuss all possible types of
solutions which  give finite contributions to the
effective gravitational constant and  which can describe
the part of the $D$--dimensional space--time on one  side
of the $d$--dimensional brane. 
\begin{itemize} 
\item 
Type A\\  
{}For each $d$ and for each $y_L$ there are solutions given by 
eq.\ (\ref{as4}) for which  $\lim_{y\to y_L}A'=\pm\infty$  
Asymptotically we have
\begin{equation}
(d-2)A'-\phi'=\frac{3-2c_i}{y-y_0}
\,.
\end{equation}
Since for $d>2$ $c_i<3/2$ for all solutions of eq.\ (\ref{ci}) 
therefore the integral (\ref{effG}) is finite and 
these solutions have finite gravitational constant 
(the corresponding solutions have singularities of the type discussed in 
the lowest order  theory in \cite{KSS}).
We can use a part of any of those solutions for $y\in(y_L, y_0)$ to
describe the space--time between the brane located at $y=y_0$ and
the singularity located at $y=y_L<y_0$ (the singularity is located 
to the left from the brane). 
The metric is bounded from above over the
whole interval ($y_L,y_0$). The length of this interval itself is also
finite. Thus, the
contribution to the effective gravitational coupling coming from such
solution is finite.

The above discussed warped metric was obtained by using
solutions with  $A'>0$ and can describe part of the
space--time ``to the left''  ($y<y_0$) from the brane. The
analogous metric for ($y>y_0$) can be  obtained in the
same way from solutions with $A'<0$ which are related by the
symmetries  (\ref{symm1}), (\ref{symm2}).

\item 
Type B\\  
{}For each $d$ and for each $y_s$ there are eight solutions
with infinite $|A''(y_s)|$ but finite values of $A'(y_s)$,
four with positive and four with negative $A'$ (they are denoted by
circles on figures 1--3). For half of them the solution
is defined for $y>y_s$ and we  can use part of such a
solution for $y\in(y_L=y_s,y_0)$ to describe the  space--time
between the singularity and the brane if $y_s<y_0$.  
Other solutions are defined for $y<y_s$  and
their part for $y\in(y_0,y_R=y_s)$ can be used when the
singularity is  to the right from the brane.

All these solutions have singularities at $y=y_s$ but of
different  nature than the solutions of type A. Now the
metric is non vanishing at $y_s$  but the curvature is singular.
We assume that such a singularity  can also be used to cut
off the space--time.

\item 
Type C\\  
Now we check whether there are solutions without any singularities
which nevertheless give finite contributions to the effective
cosmological constant. Any of such solutions must be well defined 
for all values of $y>y_0$ (or $y<y_0$). 
Let us start with solutions which approach asymptotically any of the
fixed points for $y\to\infty$ ($y\to-\infty$). Their asymptotic
behavior described by eqs.\ (\ref{as1})--(\ref{b3a3})
is enough to determine whether they can produce a finite gravitational
constant.

The solution (\ref{as1}) is not appropriate because $A'$ and
$\phi'$ tend to zero and the warp factor approaches a constant for
$y\to\pm\infty$. The integral over $y\in(y_0,\infty)$ must be
infinite.

The solutions described by eq.\ (\ref{as2}) also give infinite
contributions to the effective gravitational constant. 
We have asymptotically 
\begin{equation}
(d-2)A'-\phi'\to \sgn(y)\frac{\si}{\sqrt{\al}}
\,.
\end{equation}
After integrating over $y$ we see that this expression diverges
when $|y|\to\infty$ for both types of solutions: those which
approach a fixed point for $y\to+\infty$ as well as for those
approaching a fixed point for $y\to-\infty$.

Let us now consider the solutions described by eq.\ (\ref{as3}) for
which 
\begin{equation}
(d-2)A'-\phi'\to 
-\frac{\eps_a\ga+\eps_b\sqrt{(d-1)\ga^2-\frac12}}{\sqrt{\al}}
\end{equation}
as $y\to\pm\infty$. Using eqs.\ (\ref{c}) and (\ref{gammapm})
one can see that the sings of $(d-2)A'-\phi'$ and $c$ are the same
for $\ga=\ga_+$ while they are opposite for $\ga=\ga_-$. This means
that the solutions approaching the fixed points with
$A'=\pm\ga_-/\sqrt{\al}$ can have finite effective gravitational
coupling in $d$ dimensions and those with $A'=\pm\ga_+/\sqrt{\al}$
can not.

After analyzing which solutions approaching asymptotically the 
fixed points in $y=\pm\infty$ lead to a finite effective
gravitational constant in $d$ dimensions 
we have to analyze the isolated fixed points. The 
one with vanishing $A'$ and
$\phi'$ (given by eq.\ (\ref{FP0}) for $\La=0$) is not appropriate
because it describes just a Minkowski space with a constant metric. 
The situation is more interesting for the two fixed points of the form
(\ref{FP0}) for negative bulk cosmological constant $\La$. The warp
factor in the string frame is still constant ($A'=0$) but the dilaton is not.
We get
\begin{equation}
\int\d y\exp[(d-2)A(y)-\phi(y)]
=
\exp\left[(d-2)A_0-\phi_0\right]
\int\d y\exp
\left[-\eps\frac{\si}{\sqrt{\al}} y\right].
\end{equation}
for $\phi'=\eps\si$ with $\si$ given by eq.\ (\ref{sigma}). 
This contribution is finite
for the fixed point solution with positive $\phi'=\si$ in the region
$y\in(y_0,\infty)$ and for the fixed point solution with negative
$\phi'=-\si$ in the region $y\in(-\infty,y_0)$. 
Thus the fixed point solution $A'=0$,
$\phi'=\si$ describes a Randall--Sundrum type warped infinite
half--space to the right from the brane while that with $A'=0$,
$\phi'=-\si$ the one to the left from the brane.

The situation is similar with the fixed points with non zero $A'$
given by eqs.\ (\ref{FPnon0}) with $\ga=\ga_+$.
The contribution to the gravitational constant is equal to 
\begin{equation}
\exp\left[(d-2)A_0-\phi_0\right]
\int\d y\exp\left[
-\left(\frac{\eps_a\ga+\eps_b\sqrt{(d-1)\ga^2-\frac12}}{\sqrt{\al}}
\right)y\right].
\end{equation}
Using eqs.\ (\ref{c}) and (\ref{gammapm}) it is easy to check that the 
sign of the expression in the round
bracket above is the same as the sign of $\eps_b$.
Thus, the fixed point solutions (\ref{FPnon0}) with $\eps_b=+1$ 
($\eps_b=-1$) give finite gravitational coupling after integrating
over the region $y\in(y_0,+\infty)$ ($y\in(-\infty,y_0)$). These fixed 
points are isolated since these conditions on $\eps_b$ are just the 
opposite to the conditions for the existence of asymptotic solutions.

For $\ga=\ga_-$ fixed points with finite 
gravitational constant are always accompanied by asymptotic solutions since 
(as we have shown) the conditions on $\eps_a$ coincide with conditions 
for the existence of asymptotic solutions.
\end{itemize}

We have shown in this section that there are three types of
half--spaces 
which give finite contributions to the effective $d$--dimensional
gravitational coupling. For two of them the space--time ends at
singularities with divergent curvature. They have different behavior
of the metric at the singularity, it vanishes for type A while it is
non zero for type B solutions. Type C solutions have no singularities and 
extend to infinity.

All of the above solutions can describe one side of the 
$D$--dimensional space--time. Taking two solutions we can construct a 
full space--time on both sides of the brane. There are 
many types of such warped space--times  depending on the
solutions used to build them. We can denote them as: 
A-0-A, B-0-B or C-0-C  where the 0 in X-0-X denotes a
brane (which always can be put at $y=0$) between two solutions of type
X (one could consider also non symmetric spaces like e.g. A-0-C).  
The spaces of
type A-0-A are similar to those considered in ref.\ 
\cite{KSS} in the lowest order theory. Spaces of type B-0-B
also end at two singularities
but the character of those  singularities is different.
The curvature becomes singular like in the  A-0-A case but
the metric itself is nonsingular.  In case C-0-C the
space--time is infinite and warped in such a way that  the
$d$--dimensional effective gravitational constant is
finite (similarly to the space of the Randall--Sundrum type).

Fixed point solutions of the C-0-C type can be written explicitly. Those
leading to a finite gravitational coupling are given by:
\begin{equation}
A'=0
\,,
\qquad
\phi'=\sgn(y)\frac{\si}{\sqrt{\al}}
\,,
\label{C0C0}
\end{equation}
or
\begin{equation}
A'=-\sgn(y)\frac{\ga_+}{\sqrt{\al}}
\,,
\qquad
\phi'
=
-\sgn(y)\left(
\frac{(d-1)\ga_+-\sqrt{(d-1)\ga_+^2-\frac12}}{\sqrt{\al}}\right),
\label{C0Cnon01}
\end{equation}
or                
\begin{equation}
A'=\sgn(y)\frac{\ga_-}{\sqrt{\al}}
\,,
\qquad
\phi'
=
\sgn(y)\left(
\frac{(d-1)\ga_--\sqrt{(d-1)\ga_-^2-\frac12}}{\sqrt{\al}}\right),
\label{C0Cnon02}
\end{equation}
In all cases the brane is located at $y=0$. There are also 
solutions that asymptotically behave as (\ref{as3}) with $\ga=\ga_-$ 
and $\eps_a=\sgn(y)$.

Another interesting type of solutions with finite effective
gravitational constant exist for all dimensions $d$ different from 9.
In figures 1 and 3 one can see solutions which end at two
singularities of type B. They describe smooth space--times which have
finite length along 
the $y$ coordinate without any brane between the singular end
points. Such smooth (between the end points) brane--less
solutions can be described in our 
notation as B-B. There is also a brane--less solutions for $d=9$ but
it has slightly different character. It can be denoted as A-B model
because it has a type A singularity at one end
a type B singularity at the other end.

One should stress that most of the above solutions appear 
due to the terms of higher order in $\al'$. 
In the lowest order dilaton gravity
only solutions of the type A-0-A  are present. All other
types of solutions (A-B B-B, B-0-B, C-0-C and mixed X-0-Y)  
appear only after the next order corrections are taken into account.

\section{Equations of motion on the brane}

So far we were discussing only the bulk solutions away from the
brane. In this section the presence of the brane is taken into
account. We have to solve the eqs.\ (\ref{sumT}) and (\ref{sumW}) 
including $T_{\mu\nu}^{(B)}$ and $W^{(B)}$ given by (\ref{TB}) and
(\ref{WB}). A solution to the left (right) 
from the brane is denoted by $A_L(y)$, $\phi_L(y)$ 
($A_R(y)$, $\phi_R(y)$). We assume that $A$ and
$\phi$ are continuous across the brane while $A'$ and
$\phi'$ have jumps described by $A'_R(y_0)-A'_L(y_0)$ and
$\phi'_R(y_0)-\phi'_L(y_0)$. Therefore we can integrate
these equations over an infinitesimally small interval in
$y$ surrounding the brane. Integrating equation $\al\beta$ we get 
\begin{eqnarray} 
\frac{e^{\phi_0}}{2}V(\phi_0)
&\!\!\!=\!\!\!&
\left.\left[
(1-d)A'_R+\phi'_R
-\al\left(\frac23(d-1)(d-2)(d-3)(A'_R)^3 
-2(d-1)(d-2){A'_R}^2\phi'_R
\right.\right.\right.
\nn\\ 
&&\left.\left.\left.\qquad\qquad\qquad\qquad\qquad
+2(d-1)A'_R(\phi'_R)^2
-\frac23(\phi'_R)^3\right)
\right]\right|_{y=y_0}
-\Big[R\to L\Big],   
\label{EoMmnB} 
\end{eqnarray} 
where $\phi_0=\phi(y_0)$. The dilaton equation gives
\begin{eqnarray} 
\frac{e^{\phi_0}}{2}V'(\phi_0)
&\!\!\!=\!\!\!&
\left.\left[
dA'_R-\phi'_R
+\al\left(\frac23 d(d-1)(d-2)(A'_R)^3 
-2d(d-1){A'_R}^2\phi'_R
\right.\right.\right.
\nn\\ 
&&
\left.\left.\left.\qquad\qquad\qquad\quad
+2dA'_R(\phi'_R)^2
-\frac23(\phi'_R)^3\right)\right]\right|_{y=y_0}
-\Big[R\to L\Big].
\label{EoMdilB} 
\end{eqnarray}
Of course ($A'_R(y_0)$, $\phi'_R(y_0)$) 
and ($A'_L(y_0)$, $\phi'_L(y_0)$) have
to independently satisfy the constraint equation
(\ref{EoMyy}).

In general it is not difficult to fulfil the above equations for
solutions which {\it are not} exactly at the fixed points. 
Any of such solutions used to describe part of the
space--time laying on one side (assume for definiteness
$y>y_0$ denoted by $R$ side) of the brane has 3 free
parameters. One of them is the value of $A'$ when
approaching the brane $A'\to A'_R(y_0)$. Of course this
$A'_R(y_0)$ cannot be totally arbitrary but can take 
values only in the range allowed for a given solution (see
figures  1--3). Specifying the type of a solution and
$A'_R(y_0)$ gives uniquely a point on a curve at which the brane is 
located (except for one type of solutions when $A'(y)$ is not a
monotonic function -- then we should specify also the sign of
$A''(y_0)$). The corresponding value of $\phi'_R(y_0)$ is not a free
parameter because it is related to $A'_R(y_0)$ by the algebraic 
condition (\ref{EoMyy}).

The remaining 2 free parameters are the constants of 
integration when we calculate $A(y)$ and $\phi(y)$ from
$A'(y)$ and  $\phi'(y)$, and correspond to the symmetry
(\ref{symm1}). Those  arbitrary shifts in $A$ and $\phi$
can be for just one solution  reabsorbed in the overall
normalization of the metric.

A pair of solutions used to build a space--time solves the
bulk equations of motion for $y\ne y_0$ and has 6 free 
parameters: 
\begin{equation} 
A_L'(y_0)\,,\quad A_L(y_0)\,,\quad \phi_L(y_0)\,,\quad 
A_R'(y_0)\,,\quad A_R(y_0)\,,\quad \phi_R(y_0)\,. 
\label{DoF} 
\end{equation}

The first two requirements, that $A(y)$ and $\phi(y)$
should  be continuous at the brane, can be used to fix 
$A_R(y_0)$ and $\phi_R(y_0)$: 
\begin{equation} 
A_R(y_0)=A_L(y_0)\,,\qquad \phi_R(y_0)=\phi_L(y_0)\,. 
\end{equation} 
The remaining two constants of integration, $A_L(y_0)$
and  $\phi_L(y_0)$, are just the overall normalization of
the metric (but observe that one of them, $\phi_0$, appears on the
left hand sides of the brane equations of motion
(\ref{EoMmnB}), (\ref{EoMdilB})).  
 
Thus, we are left with two important parameters: 
$A_L'(y_0)$ and $A_R'(y_0)$. These two 
remaining degrees of freedom are exactly what we need to
solve the  equations of motion at the brane even for quite general
and independent values of the left hand sides of eqs. (\ref{EoMmnB})
and (\ref{EoMdilB}).\footnote{
In practice the use of the degrees of freedom
(\ref{DoF}) can be slightly different, because the l.h.s.s of
eqs.\ (\ref{EoMmnB}) and (\ref{EoMdilB}) do depend on the value of
$\phi(y_0)$, but number of free parameters is enough to fulfil the
brane equations of motion for a general potential $V(\phi)$.
}

The situation is quite different for the solutions which {\it are}
exactly the fixed point solutions. For them $A'$ and $\phi'$ are fixed
and constant. So a pair of such solution has only 4 free parameters
\begin{equation} 
A_L(y_0)\,,\quad \phi_L(y_0)\,,\quad A_R(y_0)\,,\quad \phi_R(y_0) 
\,.
\label{DoFfixed} 
\end{equation} 
Two of them are determined from the continuity of $A(y)$ and $\phi(y)$
across the brane while the two remaining can be reabsorbed in the
metric normalization. In this case it is clear that $V(\phi(y_0))$ and
$V'(\phi(y_0))$ can not be independent because the right hand sides of
eqs.\ (\ref{EoMmnB}) and (\ref{EoMdilB}) depend only on the values of
$A'$ and $\phi'$ which are fixed for the fixed point solutions.

Let us now check what is the relation between $V(\phi_0)$ and
$V'(\phi_0)$. Adding eqs.\ (\ref{EoMmnB}) and (\ref{EoMdilB})
we get
\begin{eqnarray}
\frac{e^{\phi_0}}{2}\left(V(\phi_0)+V'(\phi_0)\right)
&\!\!\!\!=\!\!\!\!&
\left.\left\{
A'_R
\left[
1
+2\al\left((d-1)(d-2)(A'_R)^2 
-2(d-1){A'_R}\phi'_R
+(\phi'_R)^2
\right)\right]\right\}\right|_{y=y_0}
\nn\\[2pt]
&&
-\Big\{R\to L\Big\}. 
\label{VVp}
\end{eqnarray}
The expression in the curly bracket above is nothing else as the
expression in the curly bracket in eq.\ (\ref{EoMconst}). In section 
\ref{FixedPoints} we shown that it vanishes for every fixed point
solution. Thus the brane equations for fixed point solutions are
satisfied only when
\begin{equation}
V'(\phi_0)=-V(\phi_0)
\,.
\label{VpV}
\end{equation}
This means that there are two possibilities to build brane
modes using the fixed point solutions. One is to use a potential which
satisfies eq.\ (\ref{VpV}) for arbitrary value of $\phi_0$ which means
that the potential at the brane has the form
\begin{equation}
V(\phi)=\la e^{-\phi}
\label{V}
\end{equation}
where $\la$ can be interpreted as the brane cosmological constant.
The fact that in this case the dilaton equation of motion is  
exactly equal to the potential on the brane (independently of the actual 
value of the dilaton on the brane ) was used in section 
2 to argue that with such a potential the total 
effective cosmological constant vanishes. The other possibility to solve 
the equations of motion on the brane is to 
use more general form of $V(\phi)$ and arrange the parameters in such a 
way that (\ref{VpV}) is fulfilled just at $\phi=\phi_0$. 

For solutions that asymptotically behave as (\ref{as3}) with $\ga=\ga_-$ and
$\eps_a=\sgn(y)$ the r.h.s. of (\ref{VVp}) does not vanish and can be 
changed 
by changing location of the brane. Therefore we can arrange it to be 
equal to the l.h.s. and then the equations of motion are satisfied  for 
larger class of brane potentials.

\section{Fine--tuning versus (potential) self--tuning of solutions}

In section \ref{finiteG} we have shown that in the theory defined
by the action (\ref{action}) there are several classes of solutions
which effectively in $d$--dimensions are flat and 
have finite gravitational constant. It is very interesting to check
whether those solutions require fine tuning of the parameters. 

Let us start with solutions which describe a finite interval in
the $D$--th direction ending at some singularities at $y=y_L,y_R$ and
with a brane located at $y=y_0$. The discussion of
the previous section shows that such solutions do not need any fine
tuning {\it at the brane} even for a general form of the
potential $V(\phi)$. A pair of solutions used to build a space--time
with a brane have enough free parameters to fulfil the brane
equations of motion (\ref{EoMmnB}), (\ref{EoMdilB}) for some range of
values of $V(\phi_0)$ and $V'(\phi_0)$. This is similar to the so
called self--tuning solutions discussed for the lowest order gravity
with dilaton in ref.\ \cite{KSS}.

The absence of fine tuning at the (central) brane does not
unfortunately mean that such solutions are fully satisfactory. It was
argued in ref.\ \cite{FLLN} that one should take into account also
some end point branes necessary to resolve the
singularities. Typically the solutions have not enough free parameters
and the parameters of the end point branes must be fine tuned. As a
result the 
whole solution is no longer fine--tuning free. It seems that this kind
of argumentation can be applied also to the solutions with
singularities presented in this work for the higher order theory.

Let us now discuss those solutions which have no singularities (C-0-C
in the notation of section \ref{finiteG}). We have shown in the
previous section that the equations of motion at the brane
(\ref{EoMmnB}), (\ref{EoMdilB}) can be
satisfied for fixed point solutions only when the values of the potential and
its derivative at the brane are opposite (see eq.\ (\ref{VpV})) (for 
special solutions this condition can be relaxed). 
Does this mean that fine tuning is needed for such solutions? The answer
to this question depends on the definition of fine tuning to be used.
Let us look at this problem in more detail.

The problem of fine tuning in the domain wall type solutions in
different models has been discussed recently in \cite{NoGo1,NoGo2}. 
The fine
tuning in these papers was defined as being just equivalent to the
presence of any relation between values of $V$ and $V'$ at the
brane. Some no--go theorems have been given in those papers showing
that in some theories it is impossible to avoid the fine tuning in the
brane type solutions with finite effective gravitational coupling and
without singularities. Our results are in agreement with the
spirit\footnote{
We write with ``spirit'' and not with the theorems themselves because
none of those papers considered the theory discussed in the present
paper.}
of those no--go theorems: $V(\phi_0)$ and $V'(\phi_0)$ are correlated
for solutions with finite gravitational constant and without
singularities. But we do not agree with the interpretation that this
means the presence of fine tuning. Let us use the following example to
explain this. Consider a class of theories with the action given by
eqs.\ (\ref{action})--(\ref{LB}) with the potential at the brane of
the form $V(\phi)=V_0\exp(k\phi)$. Substituting this to (\ref{EoMmnB})
and (\ref{EoMdilB}) we can see that the equations of motion on the
brane can be fulfilled for the solutions of the C-0-C type only for
$k=-1$. According to the definitions from refs.\ \cite{NoGo1,NoGo2} 
this would
mean fine tuning. But $k$ should not be treated as a parameter of a
given theory but rather as a quantity belonging to its
definition. Similarly as in the bulk Lagrangian (\ref{L0})--(\ref{L1})
the same exponential factor $\exp(-\phi)$ multiplies all the terms and
we do not treat powers of it as additional free parameters. In the
case of the bulk Lagrangian this follows form the string
theory. It is natural to expect that the string theory gives the same
factor $\exp(-\phi)$ also for the brane interactions. Then the
parameter $V_0$ can be treated as the brane cosmological constant
giving the brane Lagrangian in the form
\begin{equation}
L_B=\frac{1}{2\ka^2}\sqrt{-\tilde{g}}e^{-\phi}\la\de(y)
\,.
\label{LBla}
\end{equation}

The condition $k=-1$ should not be treated as a fine tuning also for
another reason. The main feature of usual fine tuning is that a change
of one parameter must be compensated by changing (fine tuning) some
other parameters. It is not the case for the quantity $k$ in the above
example. For $k=-1$ there are solutions of desired properties (for a
range of values of parameters $\La$ and $\la$). But for $k\ne-1$
there are no such solutions for any values of other
parameters. A change of $k$ from $-$1 can not be compensated by fine
tuning of any other parameters.

Taking the above discussion into account we can write the following
statement. For arbitrary space--dimension $D$ there is a theory
defined by the action (\ref{action}) with 
the brane interactions given by eq.\ (\ref{LBla}) (equivalent to
(\ref{LB}) with $V(\phi)=\la\exp(-\phi)$). 
In such theory there are fixed point
solutions which in $(D-1)$--dimensions are flat, have finite 
gravitational coupling and have no singularities. They are given by
eqs.\ (\ref{C0C0})  (\ref{C0Cnon01}) and (\ref{C0Cnon02}). 

For different 
brane potentials it is {\it a priori} possible to have a solution of the 
asymptotic (\ref{as3}) with $\ga=\ga_-$ and $\eps_a=\sgn(y)$ but the 
solution 
exists only for a window of large negative bulk cosmological constant. 
Those solutions and the fixed point solutions do not need fine tuning in the 
following sense: The free parameters $\La$ and $\la$ are not correlated. 
(This is in contrast with the Randall--Sundrum model where the bulk
and brane cosmological constants must be 
fine tuned against each other.) For a range of
values of these parameters there is always a solution of this
type.

This way of fulfilling the equations of motion by adjusting some
integration constants instead of the parameter of the theory are
sometimes called self tuning \cite{KSS}. The better name would be
rather: {\it potential self tuning}. The reason is that the model
allows for a self tuning (contrary to situation with fine tuning) but
the self tuning mechanism itself is still missing. We still do not
know why this particular type of solutions should be favored by the
dynamics of the theory. Nevertheless this is a much more interesting
situation than that of fine tuning.

Let us now discuss stability of the (potentially) self tuning
solutions with finite gravitational constant. 
We have two classes of solutions. One of them consists of isolated fixed 
points (\ref{C0C0}) and (\ref{C0Cnon01}). The other consists of solutions 
that asymptotically approach the fixed points (\ref{FPnon0}) with 
$\eps_a=\sgn(y)$ and the fixed point solutions 
(\ref{C0Cnon02}) themselves. The isolated fixed points are built from 
the ``repulsive''
and not ``attractive'' parts of the fixed point solutions in the sense 
that one cannot perturb the solution and it must stay exactly at the fixed 
point from infinity up to the brane. When such a brane solution of the C-0-C 
type is slightly perturbed it
develops a singularity in a finite distance from the brane. This
emphasizes even more that those solutions are only {\it potentially}
self tuning ones. 

On the other hand the fixed points (\ref{FPnon0}) with $\eps_a=1$ 
($\eps_a=-1$) for $y\to\infty$ ($y\to-\infty$) are asymptotically 
approached by a solution (also with finite gravitational constant) that 
has variable $A'$ and $\phi'$ - therefore the solution can be perturbed 
and it would be a much better candidate for the self tuning solution if not 
for the fact that the solution of this type exists only for a small 
window of large negative bulk cosmological constants.

\section{Conclusions}

In the present paper we have investigated 
the higher order gravity theory coupled to the dilaton in arbitrary
number of dimensions $D$. We included in the action the terms with
four derivatives as motivated by 
the order $\al'$ string theory corrections. Equations of motion for
such theory coupled to a brane have been derived. Their solutions of
the domain wall type have been analysed in much detail. It occurs
that many more types of solutions exist in such theory as compared to
theories without dilaton or without (some of) higher order terms. 
In particular there are several different possibilities to obtain
brane models with finite effective $(D-1)$--dimensional gravitational
constant. The space in the directions perpendicular to the brane
may end with some singularities or may be infinite. There are two
types of singularities: for one type the metric is non zero at the
singularity while for the other it vanishes. There is also another new
type of solutions which has not been found in previously analysed
models. These are brane--less solutions with finite effective
gravitational constant. For those solutions the space ends with two
singularities but is smooth between them.

However the most interesting solutions are those for which the extra
direction is infinite but the effective gravitational constant is
finite due to appropriate warp factors. There are 0 to 3 such brane
models depending on the value of the bulk cosmological constant $\La$.
It has been argued \cite{NoGo1,NoGo2} that this kind of solutions in
similar models always requires the parameters of the Lagrangian to be
fine tuned. We have shown that in the theory considered in the preset work,
with all the leading string corrections, it is not necessarily the
case. Our solutions are fine--tuning free if the bulk potential has a
string motivated form $V(\phi)=\la\exp(-\phi)$ where $\la$ is the
brane cosmological constant. With strong assumptions on the bulk 
cosmological constant ($\La_0\le\La\le\La_1$) the solutions are fine--tuning 
free for larger class of brane potentials. The absence of fine tuning means 
that for arbitrary values (within some allowed range) of the bulk and
brane cosmological constants, $\La$ and $\la$, there are solutions
with desired properties. They can be obtained by appropriate choice of
some integration constants.

The method of adjusting the integration constants instead of adjusting
the Lagrangian parameters is sometimes called self tuning. But without
any mechanism favoring those specific integration constants it can be
considered only as a potential self tuning. Thus our solutions are of
the (potential) self tuning type. They exist for a range of the
Lagrangian parameters $\La$ and $\la$. We discussed also stability of
those solutions. It occurs that under
small perturbations some of them develop singularities 
in a finite distance from the brane. There is however a class of 
solutions which can be perturbed and does not develop any singularity 
- they however exist only for a small window of large negative bulk 
cosmological constant. But anyway to
consider them as a solution to the cosmological constant problem we
would need a mechanism which could explain why out of all possible
solutions those which are flat and singularity free are favored. Such a
mechanism is still missing but those solutions are nevertheless very
interesting because they at least avoid the fine tuning of the
Lagrangian parameters.

One solution similar to those discussed above was reported in 
ref.\ \cite{BCDD} when the work on the present paper was very
advanced. However it is not possible to directly compare our results
with those presented in \cite{BCDD}. The reason is that the order
$\al'$ corrections are not the same in both cases. We used those
derived in the string frame which seems to be the best one for
considering string corrections. Moreover, the authors of \cite{BCDD}
found just one class of solutions in a 5--dimensional model while in
the present paper more types of solutions are presented for arbitrary
space--time dimensions. In addition we have discussed stability of
those solutions and found also many other classes of solutions and
their asymptotic behavior.

\vspace{10mm} 
\noindent 
{\Large \bf Acknowledgments} 

\vspace{5mm}
\noindent
We would like to thank H.P. Nilles and S. F\"orste for useful
discussions.  
\\ 
Work supported in part by the European Community's Human Potential
Programme under contracts HPRN--CT--2000--00131 Quantum Spacetime,
HPRN--CT--2000--00148 Physics Across the Present Energy Frontier
and HPRN--CT--2000--00152 Supersymmetry and the Early Universe,
and by the Polish KBN grant 2 P03B 052 16.


\end{document}